\documentclass[prd,showpacs,amsmath,amssymb]{revtex4}
\usepackage{amsmath}
\usepackage{amsmath}
\usepackage{amsmath}
\usepackage{amsfonts}
\usepackage{mathrsfs}
\usepackage{pstricks}
\usepackage{color}
\usepackage{graphicx}
\usepackage{slashed}
\usepackage{amssymb}
\usepackage{enumerate}

\newcommand{\be}{\begin{equation}}
\newcommand{\ee}{\end{equation}}
\newcommand{\ba}{\begin{array}{c}}
\newcommand{\ea}{\end{array}}
\newcommand{\bqa}{\begin{eqnarray}}
\newcommand{\eqa}{\end{eqnarray}}

\makeatletter
    
    \newcommand{\Rmnum}[1]{\expandafter\@slowromancap\romannumeral #1@}
    \makeatother

\begin{document}

\bibliographystyle{unsrt}

\title{\bf  The nature of $Z_b$ states from a combined analysis of
$\Upsilon(5S)\rightarrow h_b(mP) \pi^+ \pi^-$ and
$\Upsilon(5S)\rightarrow B^{(\ast)}\bar B^{(\ast)}\pi$}

\author{Wen-Sheng Huo$^1$, Guo-Ying Chen$^{1,2}$}

\affiliation{1) Department of Physics, Xinjiang University, Urumqi
830046, China}

\affiliation{2) State Key Laboratory of Theoretical Physics,
Institute of Theoretical Physics, Chinese Academy of Sciences,
Beijing 100190, China}

\date{\today}

\begin{abstract}

With a combined analysis of data on $\Upsilon(5S)\rightarrow
h_b(1P,2P)\pi^+\pi^-$ and $\Upsilon(5S)\rightarrow B^{(\ast)}\bar
B^{(\ast)}\pi$ in an effective field theory approach, we determine
resonance parameters of $Z_b$ states in two scenarios. In one
scenario we assume that $Z_b$ states are pure molecular states,
while in the other one we assume that $Z_b$ states contain compact
components. We find that the present data favor that there should be
some compact components inside $Z_b^{(\prime)}$ associated with the
molecular components. By fitting the invariant mass spectra of
$\Upsilon(5S)\rightarrow h_b(1P,2P)\pi^+\pi^-$ and
$\Upsilon(5S)\rightarrow B^{(\ast)}\bar B^{\ast}\pi$, we determine
that the probability of finding the compact components in $Z_b$
states may be as large as about $40\%$.
\end{abstract}


\maketitle

\section{Introduction}

Two charged bottomonium-like states $Z_b^{\pm}(10610)$ and
$Z_b^{\pm}(10650)$---denoted as $Z_b$ and $Z_b^\prime$---were
discovered by the Belle Collaboration in decays
$\Upsilon(5S)\rightarrow \Upsilon (nS)\pi^+\pi^-$ for $n=1,2,$ or
$3$ and $\Upsilon(5S)\rightarrow h_b(mP)\pi^+\pi^-$ for $m=1$ or
$2$~\cite{Collaboration:2011gja,Belle:2011aa}. The masses and decay
widths averaged over the five channels are $m_{Z_b}=10607.2\pm 2.0$
MeV, $\Gamma_{Z_b}=18.4\pm 2.4$ MeV, and $m_{Z_b^\prime}=10652.2\pm
1.5$ MeV, $\Gamma_{Z_b^\prime}=11.5\pm2.2$
MeV~\cite{Agashe:2014kda}. The average masses are about $2$ MeV
above the thresholds of both $B^\ast \bar B$ and $B^\ast \bar
B^\ast$. Recently, Belle~\cite{Adachi:2012cx} also reported the
observation of these two $Z_b$ states in $\Upsilon(5S)\rightarrow
[B\bar B^\ast+c.c.] \pi$, and $\Upsilon(5S)\rightarrow B^\ast \bar
B^\ast\pi$. The discovery of the $Z_b$ states has inspired many
interesting theoretical discussions. For example, it is suggested
that these states can be molecular states of the $B\bar B^\ast+c.c.$
or $B^\ast \bar B^\ast$ meson
pairs~\cite{Bondar:2011ev,Dong:2012hc,Li:2012uc,Li:2012as,Cleven:2011gp,Cleven:2013sq,Mehen:2011yh,Mehen:2013mva}.
They are also proposed to be candidates of tetraquark
states~\cite{Ali:2011ug}. In Ref.~\cite{Swanson:2014tra,Bugg:2011jr}
the threshold enhancements are considered to be caused by cusp
effects.

Although the masses of these $Z_b$ states determined from the
experimental fits are slightly above the thresholds, one should note
that the masses are extracted by the Breit-Wigner parametrization.
As emphasized in~\cite{Cleven:2011gp,Mehen:2013mva}, if an $S$-wave
shallow bound state exists below the threshold, the amplitude should
not be parameterized in the Breit-Wigner form. Using the line shape
for a pure bound state, Ref.~\cite{Cleven:2011gp} shows that the
data on $\Upsilon(5S)\rightarrow h_b(mP)\pi^{+}\pi^{-}$ are
consistent with the bound state nature of $Z_b^{(\prime)}$.
Furthermore, the observed enhancements in $\Upsilon(5S)\rightarrow
[B\bar B^\ast+c.c.] \pi$ and $\Upsilon(5S)\rightarrow B^\ast \bar
B^\ast\pi$ by Belle are very close to the thresholds of the
$B^{(\ast)}\bar B^{(\ast)}$ systems. It is also found that the
masses of the $Z_b$ states can be below the corresponding thresholds
if these masses are extracted from data on $\Upsilon(5S)\rightarrow
B^{(\ast)}\bar B^{(\ast)}\pi$~\cite{Adachi:2012cx}.

As a fact of observations, $Z_b$ states and their analogues in the
charmonium sector
$Z_c(3900)$~\cite{Ablikim:2013mio,Liu:2013dau,Xiao:2013iha},
$Z_c(4020/4025)$~\cite{Ablikim:2013wzq,Ablikim:2013xfr} and also the
famous $X(3872)$ appear to be strongly correlated to the thresholds
of either $B^{(\ast)}$ or $D^{(\ast)}$ pairs. This feature makes it
natural to interpret these states as molecules. However, as was
pointed out in Ref.~\cite{charm1,charm2}, it is difficult to
understand the large production rates of these states in
$B$-factories, e.g. X(3872), if these states are assumed to be
loosely bound molecular states. In particular, the recent LHCb
measurement of the ratio
$R_{\psi\gamma}=\frac{\mathcal{B}(X(3872)\rightarrow\psi(2S)\gamma)}{\mathcal{B}(X(3872)\rightarrow
J/\psi\gamma)}=2.46\pm0.64\pm0.29$~\cite{Aaij:2014ala} seems not to
support a pure $D^{\ast 0}\bar D^0$ molecular interpretation of
$X(3872)$, since $R_{\psi\gamma} $ is predicted to be rather small
for a pure $D^{\ast 0}\bar D^0$ molecule~\cite{Dong:2009uf}.
Meanwhile, a compact component inside such states can compromise
both threshold phenomena and sizeable production rates. It is shown
in Ref.~\cite{Guo:2014taa,Mehen:2011ds} that the radiative decays of
$X(3872)$ are not only sensitive to long-range parts but also to
short-range parts of the wave function. The search for a
hidden-beauty counterpart of $X(3872)$, which is usually denoted as
$X_b$, is important for understanding the structure of $X(3872)$. An
effective field theory study shows that if $X(3872)$ is a molecular
bound state of $D^{\ast 0}$ and $\bar D^0$ mesons, the heavy-quark
symmetry requires the existence of molecular bound state $X_b$ of
$B^{\ast 0}\bar B^0$ with mass of $10604$~MeV~\cite{AlFiky:2005jd}.
However, there is no significant signal of $X_b$ near the threshold
of $B^{\ast 0}\bar B^0$ in $X_b\rightarrow
\pi^+\pi^-\Upsilon(1S)$~\cite{Aad:2014ama} and in $X_b\rightarrow
\omega\Upsilon(1S)$~\cite{He:2014sqj}. Ref.~\cite{Karliner:2014lta}
suggests that $X_b$ may be close in mass to the bottomonium state
$\chi_{b1}(3P)$ and mixes with it. Therefore, the experiments which
reported observing $\chi_{b1}(3P)$ might have actually discovered
$X_b$.

Obviously, more experimental data and theoretical development are
required to clarify the nature of these near threshold states. In
Ref.~\cite{Chen:2013upa} an effective field theory (EFT) approach is
proposed for the study of near threshold states (see also an
independent study in Ref.~\cite{Meng:2014ota}). In this framework
the compositeness theorem can be incorporated with a determination
of parameter $Z$ which is the probability of finding an elementary
component in the bound state, and the nature of near threshold
states can be described by the presence of both molecular and
compact components in their wavefunctions.

The main purpose of this work is to study structure of $Z_b$ states
by doing a combined analysis of data on $\Upsilon(5S)\rightarrow
h_b(mP)\pi^+\pi^-$ and $\Upsilon(5S)\rightarrow B^{(\ast)}\bar
B^{(\ast)}\pi$ within EFT approach proposed in~\cite{Chen:2013upa}.
Our work is organized as follows: in Sec. \Rmnum{2}, we recall the
EFT approach proposed in Ref.~\cite{Chen:2013upa}. In Sec.
\Rmnum{3}, we present the analysis of the $\Upsilon(5S)\rightarrow
h_b(mP)\pi^+\pi^-$ transitions and in Sec. \Rmnum{4}, the
$\Upsilon(5S)\rightarrow B^{(\ast)}\bar B^{(\ast)}\pi$. Our
numerical results are presented in Sec. \Rmnum{5}. Finally, a brief
summary is given in Sec. \Rmnum{6}.

\begin{figure}[hbt]
\begin{center}
  \includegraphics[width=10cm]{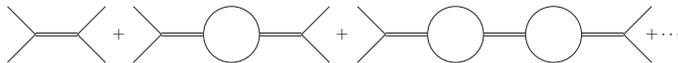}\\
  \caption{Feynman diagrams for the two particle scattering. The double lines denote the bare state.}\label{fig}
  \end{center}
\end{figure}

\section{Compositeness theorem in EFT}

In Ref.~\cite{Chen:2013upa}, we have developed an EFT approach which
incorporates Weinberg's compositeness
theorem~\cite{Weinberg1,Weinberg2}. Here we recall some of the main
points; more details can be found in Ref.~\cite{Chen:2013upa}.
Consider a bare state $|\mathcal{B}\rangle$ with bare mass $-B_0$
and coupling $g_0$ to the two-particle state, where the bare mass is
defined relative to the two-particle threshold. The two particles
have masses $m_1$, $m_2$ respectively.  If $|\mathcal{B}\rangle$ is
near the two-particle threshold, then the leading two-particle
scattering amplitude can be obtained by summing the Feynman diagrams
in Fig.~\ref{fig}. Near threshold, the momenta of these two
particles are non-relativistic. Therefore, the loop integral in
Fig.~\ref{fig} can be done in the same way as that in
Ref.~\cite{KSW1,KSW2}. The loop integral can be written as
\begin{eqnarray}
\mathcal{I}&=&\int\frac{d^D\ell}{(2\pi)^D}\frac{i}{\ell^0-\vec{\ell}^2/(2m_1)+i\epsilon}\cdot
\frac{i}{E-\ell^0-\vec{\ell}^2/(2m_2)+i\epsilon},\nonumber\\
&=&\int\frac{d^{D-1}\ell}{(2\pi)^{D-1}}\frac{i}{E-\vec{\ell}^2/(2\mu)+i\epsilon},\nonumber\\
&=&-i2\mu\Gamma(\frac{3-D}{2})(4\pi)^{\frac{1-D}{2}}(-2\mu
E-i\epsilon)^{\frac{D-3}{2}},
\end{eqnarray}
where $\mu$ is the reduced mass of the two particles, and $E$ is the
kinematic energy of the two-particle system. Obviously, the above
integral does not diverge in $D=4$. Using the minimal
subtraction(MS) scheme which subtracts the $1/(D-4)$ pole before
taking the $D\rightarrow 4$ limit, one obtains
\begin{eqnarray}
\mathcal{I}=i\frac{\mu}{2\pi}(-2\mu E-i\epsilon)^{1/2}.
\end{eqnarray}
It is interesting to note that, with the MS scheme no counter term
is needed in the renormalization. We then have the two body elastic
scattering amplitude for Fig. \ref{fig}
\begin{equation}
\mathcal{A}=-\frac{g_0^2}{E+B_0-g_0^2\frac{\mu}{2\pi}\sqrt{-2\mu
E-i\epsilon}} . \label{firstAmp}
\end{equation}
If a bound state exists, we can have the following relations
\begin{equation}
g_0^2=g^2/Z,\ \ \ \ B_0=\frac{2-Z}{Z}B, \ \ \ \
g^2=\frac{2\pi\sqrt{2\mu B}}{\mu^2}(1-Z),\label{gz}
\end{equation}
where B is the binding energy, and $Z$ is the probability of finding
an elementary state in the physical bound state. Note that for the
bound state, we mean a below threshold pole in the physical sheet.
With Eq.~(\ref{gz}), Eq.~(\ref{firstAmp}) can be re-expressed as
\begin{equation}
\mathcal{A}=-\frac{g^2}{E+B+\tilde{\Sigma}(E)},\label{AA}
\end{equation}
where
\begin{equation}
\tilde{\Sigma}(E)=-g^2[\frac{\mu}{2\pi}\sqrt{-2\mu
E-i\epsilon}+\frac{\mu\sqrt{2\mu B}}{4\pi B}(E-B)].
\end{equation}
We can also express Eq. (\ref{AA}) in the form
\begin{equation}
i \mathcal{A}=ig_0\cdot G(E)\cdot ig_0,\label{AAA}
\end{equation}
where $G(E)$ is the complete propagator for the $S$-wave near
threshold state
\begin{equation}
G(E)=\frac{iZ}{E+B+\tilde{\Sigma}(E)+i\Gamma/2}.\label{propagator}
\end{equation}
We have added a constant width $\Gamma$ in the propagator, which can
simulate the decay channels other than the bottom and anti-bottom
mesons. From Eq. (\ref{AAA}), one can find that the Feynman rule for
the coupling between the near threshold state and its two-particle
component is $ig_0$. Treating the binding momentum $\gamma=(2\mu
B)^{1/2}$ and the three-momentum of the two-particle state $p$ as
small scales, i.e., $\gamma,p\sim \mathcal{O}(p)$, one can then find
that the leading amplitude Eq. (\ref{AA}) is at the order of
$\mathcal{O}(p^{-1})$.

\begin{figure}[hbt]
\begin{center}
  \includegraphics[width=10cm]{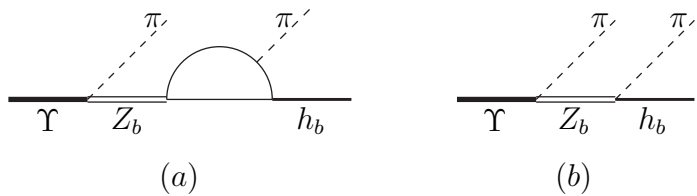}\\
  \caption{Feynman diagrams for $\Upsilon(5S)\rightarrow Z_b^{(\prime)} \pi\rightarrow h_b(mP) \pi\pi$, where the $Z_b$ states are produced in direct production processes.
  Solid lines in the loop represent bottom and anti-bottom mesons.}\label{shortfig}
\end{center}
\end{figure}

\section{$\bf \Upsilon(5S)$ decays to $\bf
{h_b(1P,2P)\pi^+\pi^-}$}\label{iii}

In this section, we study the decay $\Upsilon(5S)\rightarrow
Z_b^{(\prime)}\pi\rightarrow h_b(mP)\pi\pi$ in the EFT approach.
Generally, in the decay $\Upsilon(5S)\rightarrow Z_b^{(\prime)}\pi$,
$Z_b$ states can be produced through both direct and indirect
processes. In direct production processes, $Z_b$ states are produced
directly via its compact component, while in indirect production
processes a bottom and anti-bottom meson pair is produced first in
the $\Upsilon(5S)$ decay and then rescatters to $Z_b^{(\prime)}$.
Similarly, the decay $Z_b^{(\prime)}\rightarrow h_b(mP) \pi$ can
proceed through both direct and indirect processes. In direct decay,
$Z_b$ state will decay to $h_b(mP) \pi$ directly. In indirect decay,
$Z_b^{(\prime)}$ will first decay into a bottom and anti-bottom
meson pair and then the meson pair rescatters into $h_b(mP) \pi$.

The three-momenta of heavy mesons in decay $\Upsilon(5S)\rightarrow
Z_b^{(\prime)}\pi\rightarrow h_b(mP)\pi\pi$ are small compared with
their masses. Therefore, these heavy mesons can be treated as
non-relativistic, and one can set up a power counting in terms of
the small three-momentum
$p$~\cite{Chen:2013upa,Guo:2010ak,Guo:2009wr,Guo:2010zk}. From the
power counting, one can find that if $Z_b^{(\prime)}$ contains a
compact component, its production will be driven by this compact
component~\cite{Chen:2013upa} (see also Ref.~\cite{charm1,charm2}).
In Fig.\ref{shortfig}, we show Feynman diagrams where the $Z_b$
states are produced via compact components and decay through both
direct and indirect processes.

\begin{figure}[hbt]
\begin{center}
  \includegraphics[width=12cm]{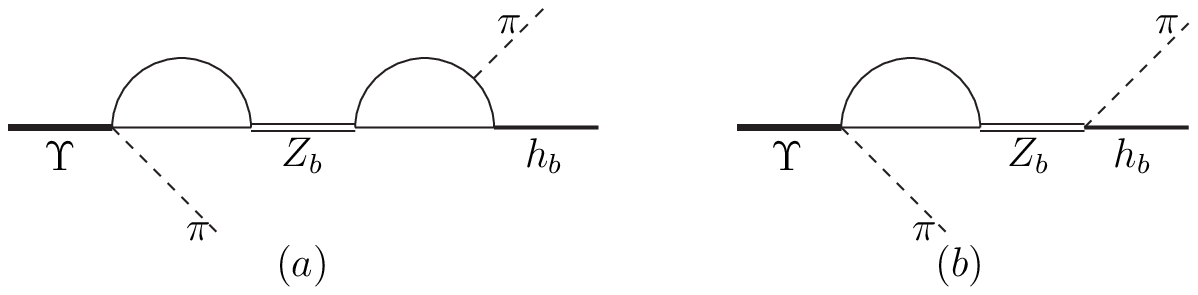}\\
  \includegraphics[width=12cm]{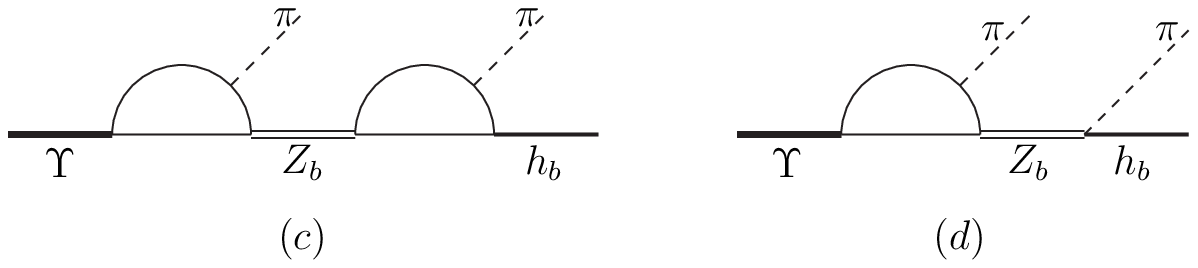}
  \caption{Feynman diagrams for $\Upsilon(5S)\rightarrow Z_b^{(\prime)} \pi\rightarrow h_b(mP) \pi\pi$, where the $Z_b$ states are produced in indirect production processes.
  Solid lines in the loops represent bottom and anti-bottom mesons.}\label{longfig}
  \end{center}
\end{figure}

If $Z_b^{(\prime)}$ is a pure molecular state, its production should
via indirect processes. The leading Feynman diagrams for indirect
production of $Z_b^{(\prime)}$ are shown in Fig.~\ref{longfig}. Note
that there are two kinds of indirect production mechanisms for $Z_b$
states. In Fig.~\ref{longfig}(a,b), $\Upsilon(5S)$ first decays to a
bottom and anti-bottom meson pair and pion in the same vertex, then
the bottom and anti-bottom meson pair rescatters to
$Z_b^{(\prime)}$. While in Fig.~\ref{longfig}(c,d), $\Upsilon(5S)$
first decays to a bottom and anti-bottom meson pair, and after
emitting one pion, the bottom and anti-bottom meson pair rescatters
to $Z_b^{(\prime)}$. It is shown in Ref.~\cite{Mehen:2013mva} that
both mechanisms contribute at leading order for the indirect
production of $Z_b$ states.

As we are only interested in low energy physics, it is convenient to
collect $B$ mesons in a $2\times 2$
matrix~\cite{Hu:2005gf,Fleming:2008yn}
\begin{equation}
H_a=\vec{P}_{a}^\ast\cdot\vec{\sigma}+P_a,\ \ \ \ \ \bar
H_a=-\vec{\bar P}_a^\ast\cdot\vec{\sigma}+\bar P_a,\ \ \ \ \ \ \
P_a^{(\ast)}=(B^{(\ast)- },\ \ \ \bar B^{(\ast)0}),
\end{equation}
where $\sigma^i$ are the Pauli matrices, and $a$ is the light flavor
index, $P_a^\ast$ and $P_a$ annihilate the vector and pseudoscalar
heavy mesons respectively, and $\bar P_a^{(\ast)}$ annihilates the
corresponding anti-particle. The leading effective Lagrangian
describing the coupling of $Z_b$ states to the bottom and
anti-bottom mesons can be written as that in
Ref.~\cite{Cleven:2011gp}
\begin{equation}
\mathcal{L}_{Z_b HH}=\frac{g_0}{2\sqrt{2}}\mbox{Tr}[Z_{ab}^{\dagger
i}H_a\sigma^i \bar
H_b]+\frac{g_0}{2\sqrt{2}}\mbox{Tr}[(Z^T)^i_{ba}\bar
H_b^\dagger\sigma^i H_a^\dagger],\ \ \ \ \ Z_{ab}=\left(
           \begin{array}{cc}
             \frac{1}{\sqrt{2}}Z_b^{(\prime)0} & Z_b^{(\prime)-} \\
             Z_b^{(\prime)+} & -\frac{1}{\sqrt{2}}Z_b^{(\prime)0} \\
           \end{array}
         \right)_{ab},\label{ZbHH}
\end{equation}
where $Z_{ab}$ annihilates $Z_{ab}$, $Z_{ab}^\dagger$ creates
$Z_{ab}$, and $g_0$ is defined in Eq. (\ref{gz}). The Lagrangian for
the coupling of the $P$-wave quarkonia and the $B$ mesons
reads~\cite{Guo:2010ak}
\begin{equation}
\mathcal{L}_{h_b HH}=\frac{g_h}{2}\mbox{Tr}[h_b^{\dagger i} H_a
\sigma^i\bar H_a]+\mbox{H.c.}\label{hbHH}
\end{equation}
The chiral Lagrangian for the $B$ mesons and the $S$-wave quarkonia
can be written as~\cite{Mehen:2013mva}
\begin{equation}
\mathcal{L}_{{\small\mbox{HH}}\chi{\small\mbox{PT}}}=g_{\pi}\mbox{Tr}[\bar
H_a^\dagger\sigma^i\bar H_b]A^i_{ab}-g_{\pi}\mbox{Tr}[H_a^\dagger
H_b\sigma^i]A^i_{ba}+\frac{1}{2}g_{1}\mbox{Tr}[\Upsilon \bar
H_a^\dagger
H_b^\dagger]A^0_{ab}+\frac{1}{2}ig_2\mbox{Tr}[\Upsilon\bar
H_a^\dagger\sigma\cdot \overleftrightarrow{\partial}
H_a^\dagger]+\mbox{H.c.},
\end{equation}
where $A\overleftrightarrow{\partial}B\equiv A(\partial B)-(\partial
A)B$, $\Upsilon$ is the $2\times 2$ matrix field defined as
$\Upsilon=\vec{\Upsilon}(5S)\cdot\vec{\sigma}+\eta_b(5S)$, and
$A_\mu$ is the axial vector pion current which is given by
\begin{equation}
A_\mu=\frac{i}{2}(\xi^\dagger\partial_\mu\xi-\xi\partial_\mu\xi^\dagger)=-\partial_\mu
M/F_\pi+\cdots,\ \ \ \ \xi=e^{\frac{iM}{F_\pi}},\ \ \ \ M=\left(
                                            \begin{array}{cc}
                                              \frac{1}{\sqrt{2}}\pi^0 & \pi^+ \\
                                              \pi^- & -\frac{1}{\sqrt{2}}\pi^0 \\
                                            \end{array}
                                          \right),\ \ \ \ \
                                          F_\pi=132\ \mbox{MeV}.
\end{equation}
We set $g_\pi=0.25$ as in~\cite{Cleven:2011gp,Li:2010rh}. Note that
our convention is different from that in~\cite{Cleven:2011gp},
because a factor of $\sqrt{2M}$ has been absorbed into the field
operator of the heavy meson in our convention~\cite{Chen:2013upa},
then our $g_\pi$ is half of the value which is used in
Ref.~\cite{Cleven:2011gp}. The leading effective Lagrangian
describing the $Z_b h_b\pi$ interactions reads
\begin{equation}
\mathcal{L}_{Z_b h_b\pi}=g_z\varepsilon^{ijk}Z_{ab}^i h_b^{ \dagger
j}A_{ab}^k+\mbox{H.c.},
\end{equation}
which describes the direct decay of $Z_b^{(\prime)}\rightarrow
h_b(mP) \pi$.

Finally, we come to the vertex describing decay of $\Upsilon(5S)$
into $Z_b^{(\prime)}\pi$. The corresponding Lagrangian to the
leading order of the chiral expansion is given
by~\cite{Cleven:2011gp}
\begin{equation}
\mathcal{L}_{\Upsilon Z_b\pi}=g_{\Upsilon}\Upsilon^i(5S)
Z_{ba}^{\dagger i}A^0_{ab}+\mbox{H.c.}.\label{YZbpi}
\end{equation}
Similar to Ref.~\cite{Cleven:2011gp}, we use the same coupling
$g_{\Upsilon}$, $g_z$ for $Z_b$ and $Z_b^\prime$.

\begin{figure}[hbt]
\begin{center}
  \includegraphics[width=15cm]{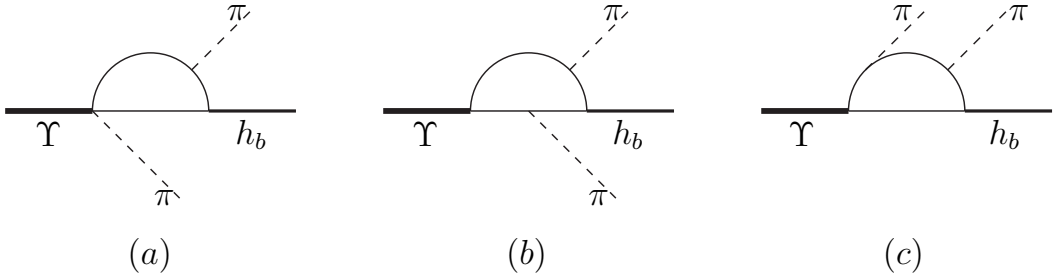}\\
  \caption{Feynman diagrams for non-resonant processes $\Upsilon(5S)\rightarrow h_b(mP) \pi \pi.$ Solid lines in the loop represent bottom and anti-bottom mesons.}\label{backfig}
  \end{center}
\end{figure}

With the above effective Lagrangians and Eq.~(\ref{propagator}) as
the propagator of $Z_b^{(\prime)}$, one can then write out the
amplitudes for all the Feynman diagrams in Fig.~\ref{shortfig}
and~\ref{longfig}. We treat the loop integrals as was done in
Ref.~\cite{Guo:2010ak}. We present the relevant one loop three-point
functions in Appendix A, and give all the amplitudes of Fig.~
\ref{shortfig} and~\ref{longfig} in Appendix B. In the following we
address several points before ending this section.

\begin{itemize}
\item As in Ref.~\cite{Cleven:2011gp}, we assume that $Z_b$ only couples to $B\bar B^\ast$ while $Z_b^\prime$
only couples to $B^\ast \bar B^\ast$. We then find that there is a
relative minus sign between $i\mathcal{M}_{3a,3b,3c,3d}$ for
$\Upsilon(5S)\rightarrow Z_b^{+} \pi^-\rightarrow h_b(mP)
\pi^+\pi^-$ and those for $\Upsilon(5S)\rightarrow Z_b^{\prime +}
\pi^-\rightarrow h_b(mP) \pi^+\pi^-$. It should not be surprising to
find this relative minus sign, since if one assumes
$Z_b$($Z_b^\prime$) couples to $B^\ast \bar B^\ast$($B\bar B^\ast$)
with the same strength as that of $Z_b(Z_b^\prime)$ couples to
$B\bar B^\ast$($B^\ast \bar B^\ast$), one would find that the meson
loop amplitudes would be suppressed in heavy-quark spin symmetry
world as noticed in~\cite{Guo:2010ak}.
\item Assuming that $Z_b$ and $Z_b^\prime$ are spin partners of each
other, we can use the same $Z$ for $Z_b$ and $Z_b^\prime$. In this
way, we can reduce the number of free parameters in our fitting.
\item We show the Feynman diagrams for non-resonant contributions to $\Upsilon(5S)\rightarrow h_b(mP)\pi\pi$
in Fig.~\ref{backfig}. Ref.~\cite{Wang:2013hga} shows that the
non-resonant diagrams do not satisfy the two-cut condition near
$\Upsilon(5S)$ region. Hence their contributions will not be
enhanced by the kinematic singularity. We do not include their
contributions in the present work, since they are suppressed by the
heavy-quark spin symmetry. The experimental fits also find no
significant non-resonant
contributions~\cite{Collaboration:2011gja,Belle:2011aa}.
\end{itemize}

\begin{figure}[hbt]
\begin{center}
\includegraphics[width=15cm]{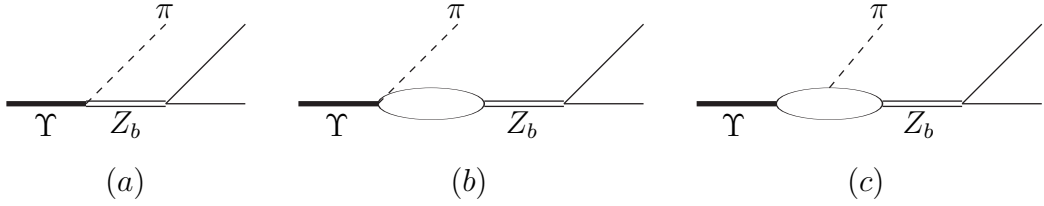}\\
\caption{Feynman diagrams for $\Upsilon(5S)\rightarrow
Z_b^{(\prime)}\pi\rightarrow B^{(\ast)}\bar B^{(\ast)} \pi.$ Solid
lines in the loop and in the final state represent bottom and
anti-bottom mesons. }\label{feyndiagram}
\end{center}
\end{figure}

\section{$\bf\Upsilon(5S)$ decays to $\bf {B^{(\ast)}\bar B^{(\ast)}\pi}$ }

In this section, we will study the decay $\Upsilon(5S)\rightarrow
B^{(\ast)}\bar B^{(\ast)}\pi$ in EFT. For the previous study one may
refer to Ref.~\cite{Mehen:2013mva}, where the $Z_b$ states are
assumed to be molecules. Instead of fitting data directly,
Ref.~\cite{Mehen:2013mva} constrains some parameters using data on
$\Upsilon(5S)\rightarrow B^{(\ast)}\bar B^{(\ast)}$,and it then
calculates the differential distribution for
$\Upsilon(5S)\rightarrow B^{(\ast)}\bar B^{(\ast)}\pi$ as a function
of invariant mass of the $B^{(\ast)}\bar B^{(\ast)}$ pair. In this
work, we give the amplitudes for $\Upsilon(5S)\rightarrow
B^{(\ast)}\bar B^{(\ast)}\pi$ in EFT and constrain parameters by
fitting the data directly.

Similar to $\Upsilon(5S)\rightarrow Z_b^{(\prime)}\pi\rightarrow
h_b(mP) \pi\pi$, $Z_b$ states can be produced through both direct
and indirect processes. The leading order Feynman diagrams for these
two different production mechanisms are presented in
Fig.~\ref{feyndiagram}. The Feynman diagrams for the non-resonant
contributions are shown in Fig.~\ref{backfig2}. We give all the
amplitudes for Fig.~\ref{feyndiagram} and~\ref{backfig2} in Appendix
C.

\begin{figure}[hbt]
  \includegraphics[width=10cm]{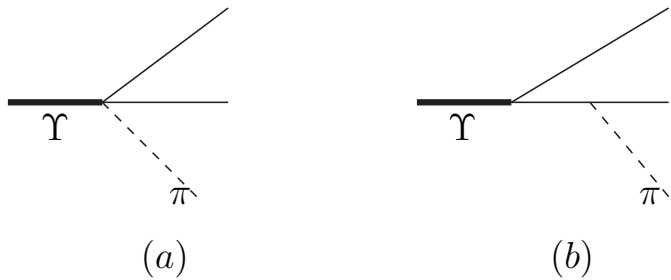}\\
  \caption{Feynman diagrams of the non-resonant contribution to $\Upsilon(5S)
\rightarrow B^{(\ast)}\bar B^{(\ast)} \pi$. The solid lines
represent the bottom and anti-bottom mesons.}\label{backfig2}
\end{figure}

\section{Numerical results }
With the amplitudes given in Appendix B and C, we do a combined fit
to data on $\Upsilon(5S)\rightarrow
h_b(mP)\pi^+\pi^-$~\cite{Collaboration:2011gja,Belle:2011aa} and
$\Upsilon(5S)\rightarrow B^{(\ast)}\bar
B^{(\ast)}\pi$~\cite{Adachi:2012cx}. Data on
$\Upsilon(5S)\rightarrow B^{(\ast)}\bar B^{\ast}\pi$ are
nonvanishing below the $B^{(\ast)}\bar B^{(\ast)}$ thresholds, hence
we have to convolve the invariant mass spectra with detector
resolution function. Data on $\Upsilon(5S)\rightarrow
h_b(mP)\pi^+\pi^-$ are collected per 10~MeV, so the invariant mass
spectra should be convolved with detector resolution function and
integrated over 10~MeV histogram bin. The detector resolution is
parameterized by a Gaussian function with energy resolution
parameter $\sigma=5.2$~MeV for $\Upsilon(5S)\rightarrow
h_b(mP)\pi^+\pi^-$~\cite{Collaboration:2011gja} and $\sigma=6$~MeV
for $\Upsilon(5S)\rightarrow B^{(\ast)}\bar
B^{(\ast)}\pi$~\cite{Adachi:2012cx}. To compare different scenarios
for the structure of $Z_b$ states, we do the fit with two
alternative schemes:

\begin{enumerate}[a.]
\item We assume that $Z_b$ states are pure molecular states, then we have to
set $Z=0$ in the fit. In this way, only the diagrams in
Fig.~\ref{longfig}(a,c),~\ref{feyndiagram}(b,c) and~\ref{backfig2}
give nonvanishing amplitudes.
\item We assume that $Z_b$ states contain substantial compact
components, i.e., tetraquark component. It is shown in
Ref.~\cite{charm2} that the production rate of a molecular state is
proportional to its wave function square at the origin
$|\Psi(0)|^2$. Because the wave function of the molecular component
in a loosely bound state spreads far out in space, $|\Psi(0)|^2$ is
quite small, then the production rate of $Z_b^{(\prime)}$ through
the molecular component will be suppressed. Therefore, we further
assume that $Z_b^{(\prime)}$ is mainly produced through the compact
component, and we set $g_1=g_2=0$ in the fitting. It is worth
mentioning that Ref.~\cite{charm1,Chen:2013upa} demonstrate that the
production of a near threshold state (by which we mean a mixture of
the compact component and molecular component) is driven by the
compact component. On the other hand, the hadronic decays of
$Z_b^{(\prime)}$ into $h_b(mP)\pi$ will mainly go through the
molecular component. This can be found from the power counting
analysis. We treat the binding momentum $\gamma$, the three momentum
of the bottom meson $p_B$ and the four momentum of pion $p_\pi$ as
small scales, i.e., they are all at the order of $\mathcal{O}(p)$.
Note that in the non-relativistic effective field theory, the
propagator of the heavy meson is at the order of
$\mathcal{O}(p^{-2})$, and the measure of the one loop integration
is at the order of $\mathcal{O}(p^5)$. One can then find that
Fig.~\ref{shortfig}(a) is at the order of $\mathcal{O}(p^{-1/2})$,
while Fig.~\ref{shortfig}(b) is at the order of $\mathcal{O}(p^0)$.
Thus, as a leading order study, we set $g_z=0$ and neglect the
contribution from Fig.~\ref{shortfig}(b). Up to now, we have shown
that while the production of $Z_b^{(\prime)}$ is driven by the
compact component, its hadronic decays mainly go through the
molecular component. It is interesting to note that similar features
are adopted for $X(3872)$ in Ref.~\cite{charm1}. By setting
$g_1=g_2=g_z=0$, one can find that only the diagrams Fig.~
\ref{shortfig}(a) and~\ref{feyndiagram}(a) give nonvanishing
amplitudes, and the number of the relevant free parameters in this
scheme is the same as that in scheme (a) (see Table.~\ref{table}).

\end{enumerate}

We then compare our fitting schemes with that used in
Ref.~\cite{Cleven:2011gp}. Although scheme(b) and
Ref.~\cite{Cleven:2011gp} use the same decay mechanism for
$\Upsilon(5S)\rightarrow Z_b\pi\rightarrow h_b\pi\pi$ as shown in
Fig.~\ref{shortfig}(a), there are some differences between them. The
main difference is that Ref.~\cite{Cleven:2011gp} sets $Z=0$, while
in scheme(b) we let $Z$ to be a free parameter which satisfies
$0<Z<1$. As shown explicitly in Appendix B, the amplitude for
Fig.~\ref{shortfig}(a) is zero by setting $Z=0$. Physically, by
setting $Z=0$, one assumes the $Z_b$ states as pure molecular states
which do not contain compact components, hence they cannot be
produced through the compact components. Therefore, if one uses
Fig.~\ref{shortfig}(a) to describe the decay mechanism of
$\Upsilon(5S)\rightarrow Z_b\pi \rightarrow h_b\pi\pi$, one cannot
set $Z=0$ as in Ref.~\cite{Cleven:2011gp}. The consistent treatment
is to let $Z$ to be a free parameter which satisfies $0<Z<1$. On the
other hand, if one assumes $Z_b^{(\prime)}$ to be a pure molecular
state, i.e., $Z=0$, one should note that it can only be produced
through indirect process. Therefore, for the pure molecular
scenario, one should use Fig.~\ref{longfig}(a,c), i.e., scheme(a),
instead of Fig.~\ref{shortfig}(a) to describe the decay mechanism of
$\Upsilon(5S)\rightarrow Z_b \pi\rightarrow h_b \pi\pi$.

Now we come to discuss the applicability of EFT. In the decay
$Z_b^{(\prime)}\rightarrow h_b(2P)\pi$, the momentum of the pion is
around $300\sim 400$~MeV in the energy region of our concern, hence
the pion can be treated as soft, and one would expect the EFT
expansion can converge fast enough. But in
$Z_b^{(\prime)}\rightarrow h_b(1P)\pi$, the momentum of the pion is
relatively large and around $600\sim 700$~MeV. Based on naive
dimensional analysis, Ref.~\cite{Cleven:2013sq} warns that the EFT
expansion may not be good enough for decay
$Z_b^{(\prime)}\rightarrow h_b(1P) \pi$ due to the relatively large
pion momentum. However, the results from the complete loop
calculations can be more complex than the naive dimensional
analysis. One may refer to Ref.~\cite{Fleming:1999ee} for an
example. Generally, it is complex to study the convergence of the
effective field theory, and reliable conclusions can only be
achieved once the complete higher loop contributions are available.
Since such a kind of study is beyond the scope of the this work, we
take a more pragmatic approach with two options in the fit.
\begin{enumerate}[1.]
\item We use data sets of $\Upsilon(5S)\rightarrow h_b(1P,2P)
\pi^+\pi^-$, $\Upsilon(5S)\rightarrow B\bar B^{\ast}\pi$ and
$\Upsilon(5S)\rightarrow B^{\ast}\bar B^{\ast}\pi$ in our fit.
\item We use data sets of $\Upsilon(5S)\rightarrow h_b(2P)
\pi^+\pi^-$, $\Upsilon(5S)\rightarrow B\bar B^{\ast}\pi$ and
$\Upsilon(5S)\rightarrow B^{\ast}\bar B^{\ast}\pi$ in our fit.
\end{enumerate}

\begin{figure}[hbt]
  \includegraphics[width=8cm]{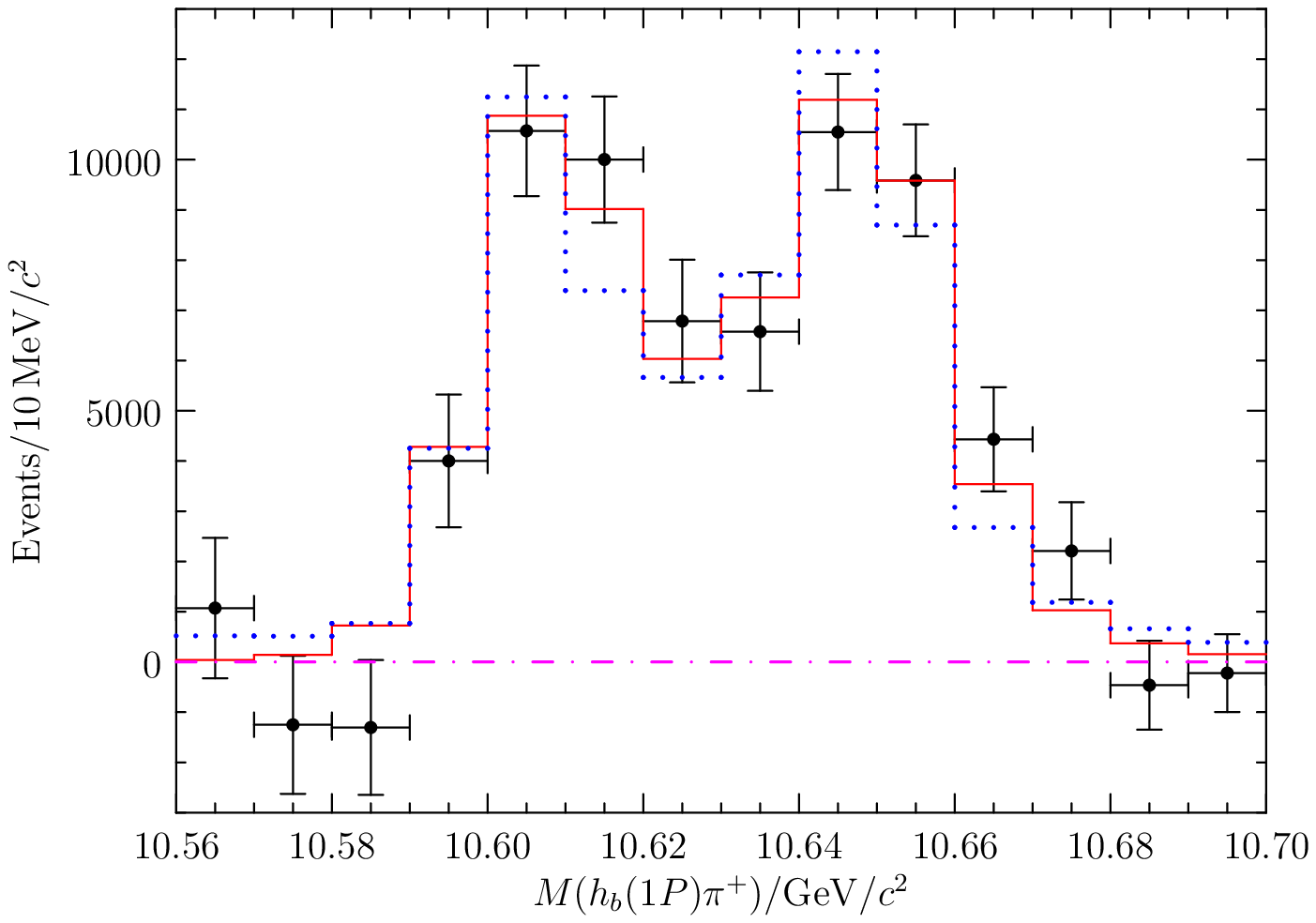}\ \ \ \includegraphics[width=8cm]{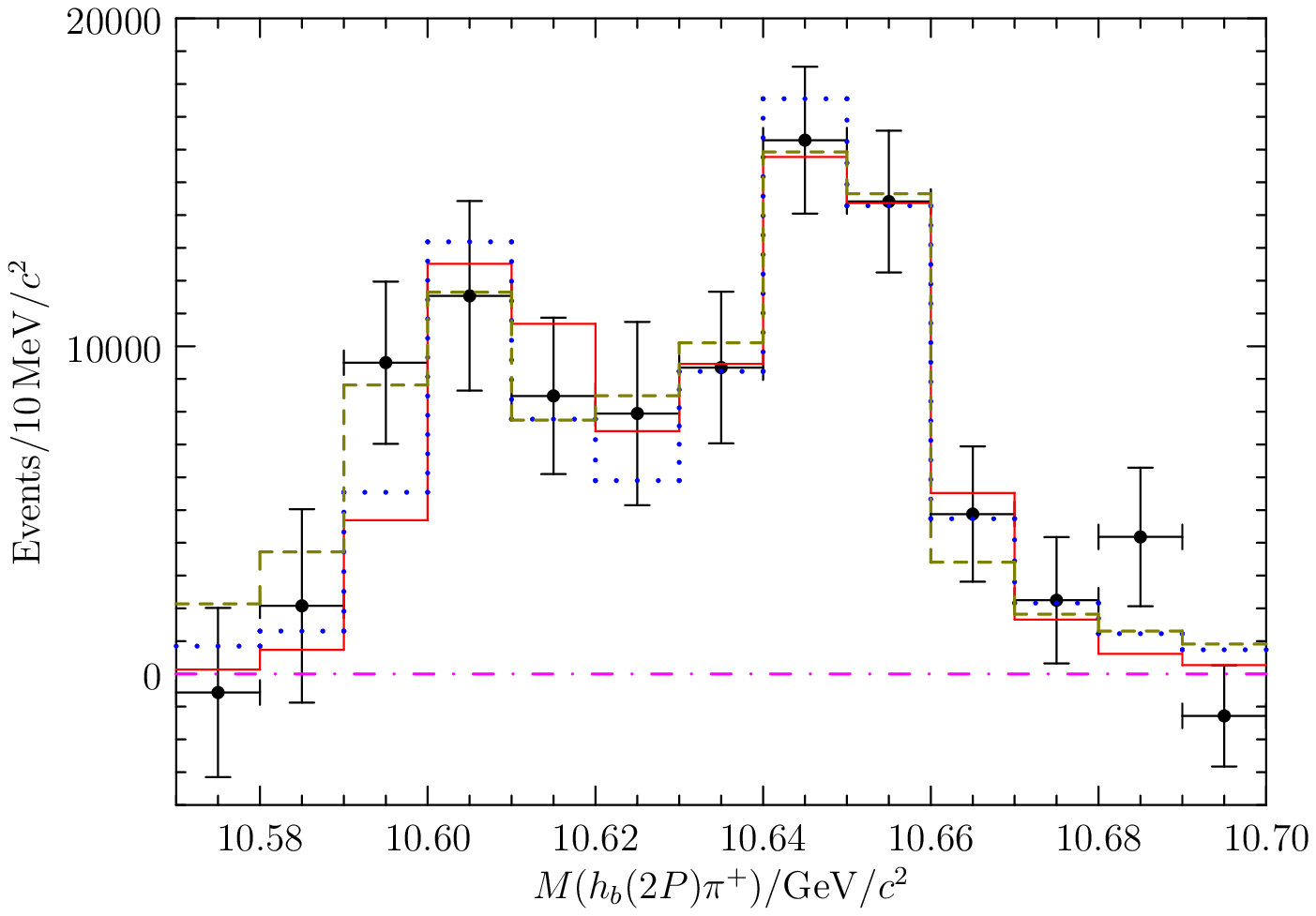}
  \caption{Comparison of the invariant mass spectra of $h_b(1P)\pi$ and $h_b(2P)\pi$
  in fit(1a), fit(2a),fit(1b) and the experiment. The dotted line is the result of fit(1a). The dashed line is the result of fit(2a). The solid line is the result of fit(1b). Data are from~\cite{Collaboration:2011gja}.}\label{hbpipi}
\end{figure}

\begin{figure}[hbt]
  \includegraphics[width=8.2cm]{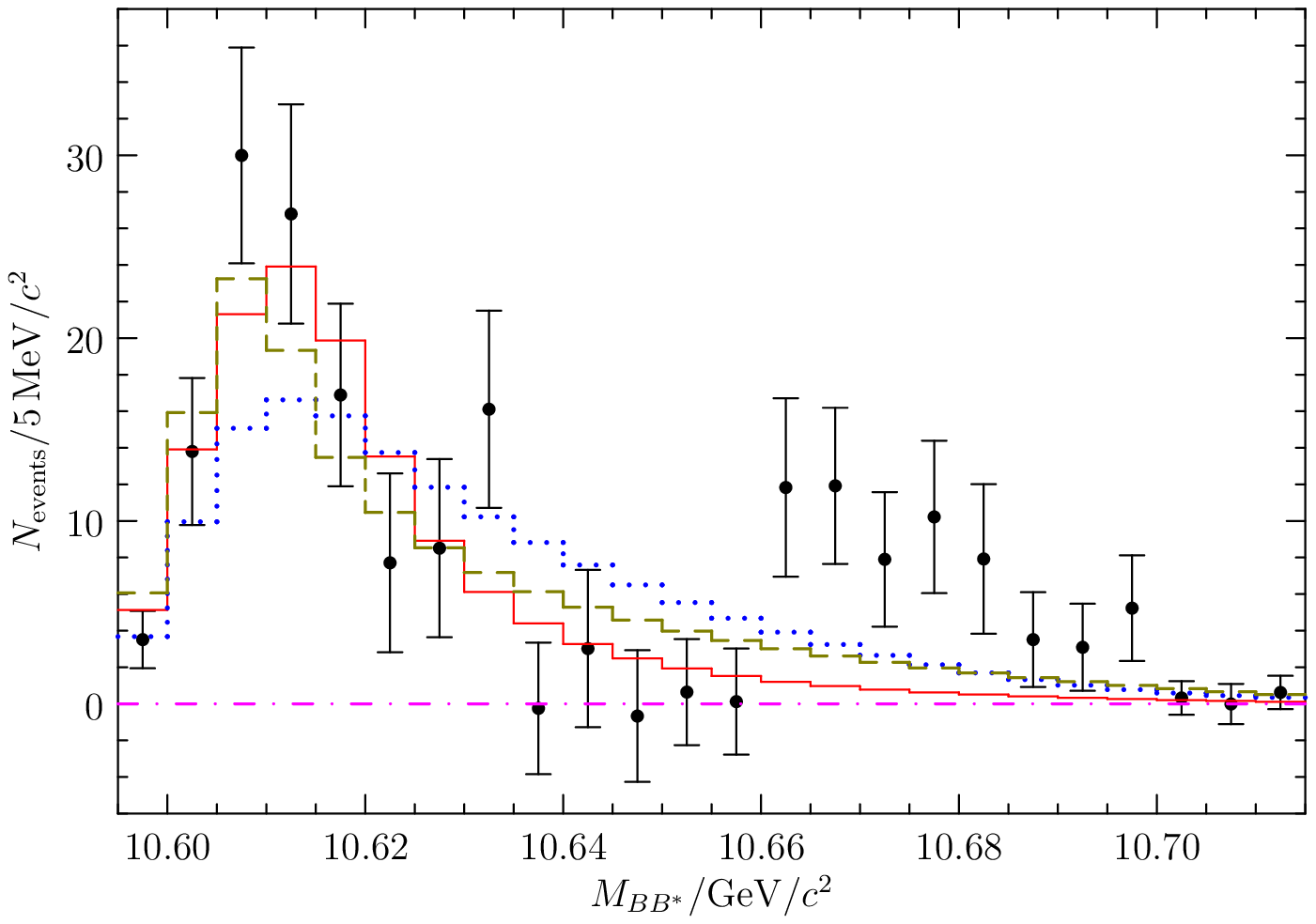}\ \ \ \includegraphics[width=8cm]{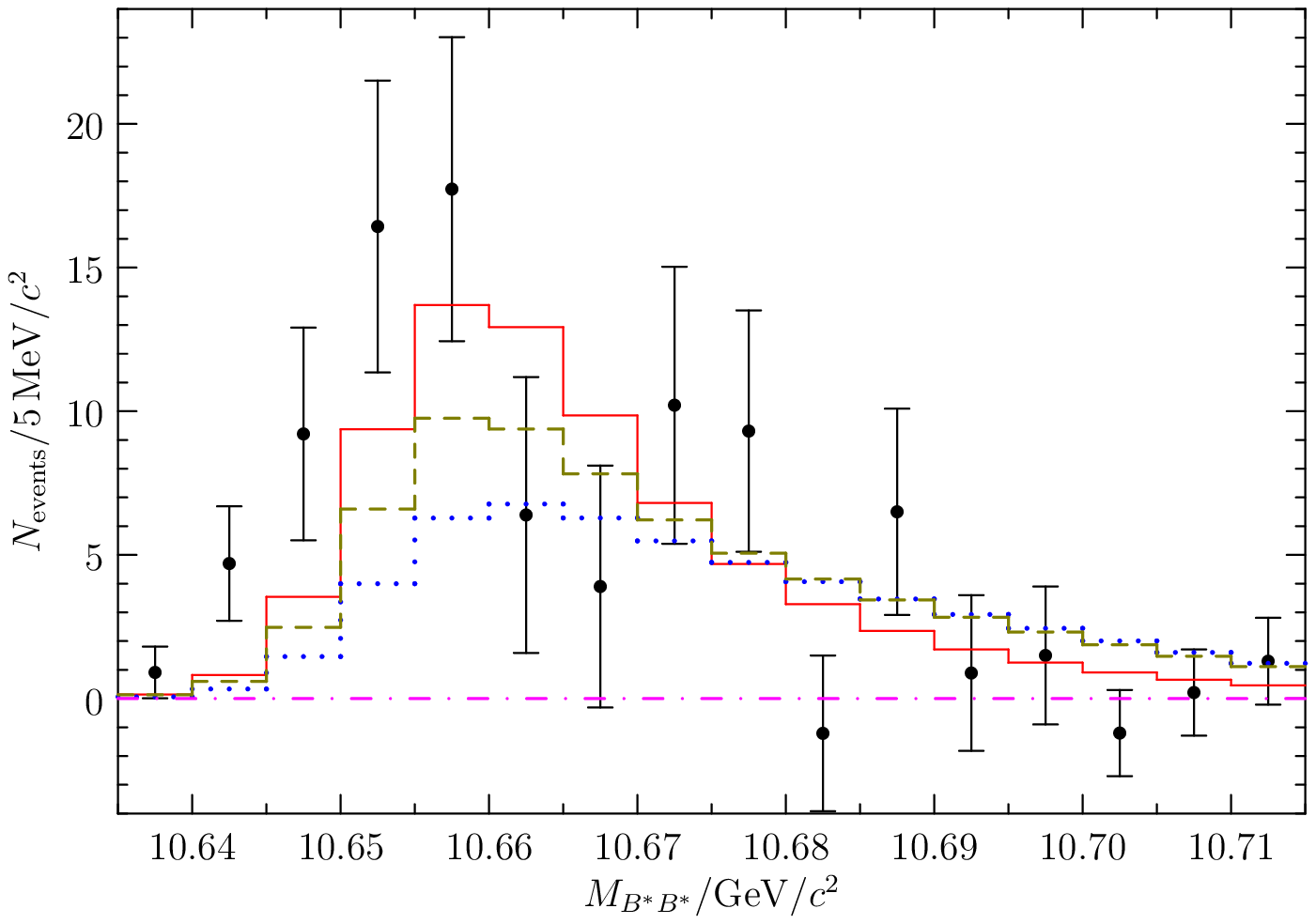}
  \caption{Comparison of the invariant mass spectra of $B\bar B^\ast $ and $B^\ast \bar B^\ast$
  in fit(1a), fit(2a), fit(1b) and the experiment. The dotted line is the result of fit(1a). The dashed line is the result of fit(2a). The solid line is the result of fit(1b). Data are from~\cite{Adachi:2012cx} and have had
  background subtracted.}\label{BBpi}
\end{figure}

We choose an individual normalization factor for each final state in
the fit. In this way, we need not to fix values of $g_{\Upsilon}$
and $g_h$. We present all the fitted parameters in
Table.~\ref{table}, and we show the fitting results of fit(1a),
fit(2a) and fit(1b) in Fig.\ref{hbpipi},~\ref{BBpi}. Note that the
width $\Gamma$ in Table.~\ref{table} is not the total width, but the
width defined in Eq.~(\ref{propagator}).

\begin{table}[!h]
\tabcolsep 0pt \caption{Parameters for four fits.} \vspace*{-12pt}
\begin{center}
\def\temptablewidth{0.8\textwidth}
{\rule{\temptablewidth}{1pt}}
\begin{tabular*}{\temptablewidth}{@{\extracolsep{\fill}}cccccccc}
Fit & $g_2/g_1$ &$B$ &$B^\prime$ &$\Gamma_{Z_b}$ &$\Gamma_{Z_b^\prime}$ &$Z$ & $\chi^2/d.o.f.$\\
\hline
       1a \ \     &   0.049(15) \ \ \ \   &\ \  0.11(12)eV\ \  \ \ &\ \  27(58)keV \ \ \ \   & 2(1)keV\ \ \ \ &  1.9(1.9)MeV\ \ \ \ & 0 & 110/58\\
       2a  \    &   0.0017(69)    &  12(21)keV   & 0.14(7)MeV & 0.12(8)MeV & 0.59(27)MeV & 0 & 72/45 \\
       1b  \ \   & --        &  0.19(22)eV  & 1.6(1.8)eV & 5.5(1.8)MeV & 7.8(2.2)MeV & 0.42(12) & 81/58 \\
       2b \ \ & -- & 0.38(65)eV & 0.51(86)eV & 6.1(2.8)MeV &4.6(2.2)MeV  &  0.42(18)& 69/45
       \end{tabular*}
       {\rule{\temptablewidth}{1pt}}
       \end{center}\label{table}
       \end{table}

We give some brief discussions as regards our fitting results as
follows
\begin{itemize}
\item
It is found in the experimental fits that the relative phase between
$Z_b$ and $Z_b^\prime$ in the $h_b(mP)\pi\pi$ channel is
$180^0$~\cite{Collaboration:2011gja,Belle:2011aa}. In fitting scheme
(a), the relative minus sign between $i\mathcal{M}_{3a,3c}$ for
$Z_b$ and $Z_b^\prime$ can account for this relative phase. However,
one can not find such a relative phase in amplitudes which are used
in scheme (b). In our fitting, we find that scheme (b) gives a good
fit only if such a relative phase is included. This may be
attributed to $g_\Upsilon$ (defined in Eq. (\ref{YZbpi})) which has
a relative minus sign between $Z_b$ and $Z_b^\prime$. We note that a
very recent paper Ref.~\cite{Ali:2014dva} proposed an explanation
for this relative minus sign.
\item From the fitting results of fit(1b) and fit(2b), one can find
that the fitted parameters in fit(1b) and fit(2b) are close to each
other. This indicates that the fitting results in scheme (b) are not
sensitive to data on $\Upsilon(5S)\rightarrow h_b(1P)\pi^+\pi^-$.
Whether this means that the effective field theory can be
successfully applied in $\Upsilon(5S)\rightarrow h_b(1P)\pi^+\pi^-$
needs to be further investigated. Nevertheless, our numerical
results show that such a possibility exists. It is also interesting
to find that in scheme (b) the fitted binding energy and the width
of $Z_b$ are close to those of $Z_b^\prime$. This seems to be
consistent with the heavy-quark spin symmetry.
\item With all data sets, scheme (1b) gives much better fitting
quality than scheme (1a). Unfortunately, if data on
$\Upsilon(5S)\rightarrow h_b(1P) \pi^+\pi^-$ are dropped, the two
schemes give almost equal fitting qualities. In this sense, it seems
too early to claim conclusively that $Z_b$ states contain
substantial compact components. However, a substantial compact
component in $Z_b^{(\prime)}$ can explain its large production rates
in experiments. In contrast, a pure molecular state with the tiny
binding energy as determined in scheme (a) is not likely to have
large production rates in $\Upsilon(5S)$ decays.
\item The binding energies of the $Z_b$ states from the fit are
generally very small. If we fix $B=0.1$~MeV, which is the case for
$X(3872)$, and $Z=0.4$ in fit(1b), we get a fitting quality
$\chi^2=90$, which is still acceptable and better than fit(1a). The
other fitting parameters are $B^\prime=0.23(14)$~MeV,
$\Gamma_{Z_b}=6.5(9)$~MeV and $\Gamma_{Z_b^\prime}=5.6(9)$~MeV. This
result also seems to be consistent with the heavy-quark symmetry.
\item One can also analysis data on $\Upsilon(5S)\rightarrow \Upsilon(nS)\pi^+\pi^-$
in the EFT approach. However, different from $h_b(mP)\pi^+\pi^-$ and
$B^{(\ast)}\bar B^{(\ast)}\pi$, the non-resonant contribution in
$\Upsilon(nS)\pi^+\pi^-$ is significant. It is impossible to
consider the interference with the non-resonant contribution
correctly in one-dimensional analysis. To analysis data on
$\Upsilon(5S)\rightarrow \Upsilon(nS)\pi^+\pi^-$, one needs to fit
the two-dimensional Dalitz distribution, which is beyond the scope
of the present manuscript.
\end{itemize}

\section{Summary}
We have done a combined analysis of data on $\Upsilon(5S)\rightarrow
h_b(1P,2P)\pi^+\pi^-$, $\Upsilon(5S)\rightarrow B\bar B^{\ast}\pi$
and $\Upsilon(5S)\rightarrow B^\ast \bar B^\ast\pi$ within EFT
approach. With a combined analysis, we determine the resonance
parameters of $Z_b$ states in two scenarios. In one scenario we
assume that $Z_b$ states are pure molecular states, while in the
other one we assume that $Z_b$ states contain compact components. It
is found that by assuming that $Z_b$ states contain substantial
compact components, one can have a better description of all data
than by pure molecular assumption. By fitting the invariant mass
spectra of $\Upsilon(5S)\rightarrow h_b(1P,2P)\pi^+\pi^-$ and
$\Upsilon(5S)\rightarrow B^{(\ast)}\bar B^{(\ast)}\pi$, we determine
that the probability of finding a compact component in
$Z_b^{(\prime)}$ is about $40\%$ . It is also interesting to note
that the probability of finding a compact component in
$Z_b^{(\prime)}$ could be close to that in $X(3872)$ which are
around $26\%\sim 44\%$~\cite{Meng:2013gga,Cardoso:2014xda}.

\section*{ACKNOWLEDGMENTS}
We would like to thank Qiang Zhao for a careful reading of the
manuscript and valuable comments and Feng-Kun Guo for very useful
discussions. We would also like to thank Guang-Yi Tang for the
helpful discussions on the Belle result. This work is supported, in
part, by National Natural Science Foundation of China (Grant Nos.
11147022 and 11305137) and Doctoral Foundation of Xinjiang
University (No. BS110104).

\section*{APPENDIX A: ONE LOOP THREE POINT FUNCTIONS}
The three-point loop functions we will encounter are
\begin{eqnarray}
&&I(m_1,m_2,m_3,q)\nonumber\\
&=&(-i)\int\frac{d^d\ell}{(2\pi)^d}\frac{1}{(\ell^0-\frac{\vec{\ell}^2}{2m_1}+i\epsilon)
(\ell^0+b_{12}+\frac{\ell^2}{2m_2}-i\epsilon)[\ell^0+b_{12}-b_{23}-\frac{(\vec{\ell}-\vec{q})^2}{2m_3}+i\epsilon]}\nonumber\\
&=&\frac{\mu_{12}\mu_{23}}{2\pi
}\frac{1}{\sqrt{a}}\left[\tan^{-1}(\frac{c^\prime-c}{2\sqrt{a(c-i\epsilon)}})+\tan^{-1}(\frac{2a+c-c^\prime}{2\sqrt{a(c^\prime-a-i\epsilon)}})\right],
\end{eqnarray}

\par

\begin{eqnarray}
&&I^{(1)}(m_1,m_2,m_3,q)\nonumber\\
&=&\frac{\mu_{12}m_3}{2\pi
\vec{q}^2}(\sqrt{c-i\epsilon}-\sqrt{c^\prime-a-i\epsilon})+\frac{m_3(c^\prime-c)}{2\mu_{23}\vec{q}^2}I(m_1,m_2,m_3,q),
\end{eqnarray}
where $\mu_{ij}=m_im_j/(m_i+m_j)$ are the reduced masses,
$b_{12}=m_1+m_2-M$, $b_{23}=m_2+m_3+q^0-M$, and
\begin{equation}
a=\left(\frac{\mu_{23}}{m_3}\right)^2\vec{q}^2,\ \ \ \ \ \
c=2\mu_{12}b_{12},\ \ \ \ \ \
c^\prime=2\mu_{23}b_{23}+\frac{\mu_{23}}{m_3}\vec{q}^2.
\end{equation}
$I^{(1)}(m_1,m_2,m_3,q)$ is defined as
\begin{equation}
q^i
I^{(1)}(m_1,m_2,m_3,q)=(-i)\int\frac{d^d\ell}{(2\pi)^d}\frac{\ell^i}{(\ell^0-\frac{\vec{\ell}^2}{2m_1}+i\epsilon)
(\ell^0+b_{12}+\frac{\ell^2}{2m_2}-i\epsilon)[\ell^0+b_{12}-b_{23}-\frac{(\vec{\ell}-\vec{q})^2}{2m_3}+i\epsilon]}\nonumber\\
.\end{equation} For more details, one can refer to
Ref.~\cite{Guo:2010ak}.

\section*{APPENDIX B: AMPLITUDES FOR $\Upsilon(5S)\rightarrow h_b(1P,2P)\pi^+\pi^-$}
The amplitudes for $\Upsilon(5S)\rightarrow Z_b^+ \pi^-\rightarrow
h_b(mP) \pi^+\pi^-$ in Fig.~\ref{shortfig} and~\ref{longfig} read
\begin{eqnarray}
i\mathcal{M}_{2a}&=&\frac{2\sqrt{2Z}g g_{\Upsilon} g_\pi g_h
E_\pi}{F_\pi^2}\frac{1}{E+B+\tilde{\Sigma}(E)+i\Gamma_{Z_b}/2}\varepsilon^{ijk}q^i\epsilon^{\ast
j}(h_b)\epsilon^{k}(\Upsilon)\nonumber\\
&&\times [I(m_{B^\ast},m_{B},m_{B^\ast},q)+I(m_B, m_{B^\ast},
m_{B^\ast},q )].
\end{eqnarray}

\begin{eqnarray}
i\mathcal{M}_{2b}&=&-i\frac{g_\Upsilon g_z
E_\pi}{F_\pi^2}\frac{Z}{E+B+\tilde{\Sigma}(E)+i\Gamma_{Z_b}/2}\varepsilon^{ijk}q^i\epsilon^{\ast
j}(h_b)\epsilon^k(\Upsilon).
\end{eqnarray}

\begin{eqnarray}
i\mathcal{M}_{3a}&=&-\frac{2g^2g_1g_\pi g_h
E_\pi}{F_\pi^2}\frac{\mu}{\pi}(-2\mu E-i\epsilon)^{1/2}\frac{1}{E+B+\tilde{\Sigma}(E)+i\Gamma_{Z_b}/2}\nonumber\\
&&\times \varepsilon^{ijk}q^i\epsilon^{\ast
j}(h_b)\epsilon^{k}(\Upsilon)
[I(m_{B^\ast},m_{B},m_{B^\ast},q)+I(m_B, m_{B^\ast}, m_{B^\ast},q
)].
\end{eqnarray}

\begin{eqnarray}
i\mathcal{M}_{3b}&=&i\frac{\sqrt{2Z}}{2\pi}\frac{gg_1g_z
E_\pi}{F_\pi^2}\mu(-2\mu
E-i\epsilon)^{1/2}\frac{1}{E+B+\tilde{\Sigma}(E)+i\Gamma_{Z_b}/2}\varepsilon^{ijk}q^i\epsilon^{\ast
j}(h_b)\epsilon^k(\Upsilon).
\end{eqnarray}

\begin{eqnarray}
i\mathcal{M}_{3c}&=&-16\frac{g^2 g_2 g_h
g_\pi^2}{F_\pi^2}\left[\left(I^{(1)}(m_{B^\ast},m_{B^\ast},m_B,p)+I^{(1)}(m_{B^\ast},m_B,m_{B^\ast},p)\right)\vec{p}^2\delta^{km}+\right. \nonumber\\
&&\left.
\left(I^{(1)}(m_{B},m_{B},m_{B^\ast},p)-I^{(1)}(m_{B^\ast},m_B,m_{B^\ast},p)\right)p^k
p^m\right]\times
\frac{1}{E+B+\tilde{\Sigma}(E)+i\Gamma_{Z_b}/2}\nonumber\\
&&\times\varepsilon^{ijk}q^i\epsilon^{\ast
j}(h_b)\epsilon^m(\Upsilon)[I(m_{B^\ast},m_B,m_{B^\ast},q)+I(m_B,m_{B^\ast},m_{B^\ast},q)].
\end{eqnarray}

\begin{eqnarray}
i\mathcal{M}_{3d}&=&i4\sqrt{2}\frac{g g_2 g_z g_\pi}{F_\pi^2}\left[\left(I^{(1)}(m_{B^\ast},m_{B^\ast},m_B,p)+I^{(1)}(m_{B^\ast},m_B,m_{B^\ast},p)\right)\vec{p}^2\delta^{km}+\right. \nonumber\\
&&\left.
\left(I^{(1)}(m_{B},m_{B},m_{B^\ast},p)-I^{(1)}(m_{B^\ast},m_B,m_{B^\ast},p)\right)p^k
p^m\right]\times
\frac{\sqrt{Z}}{E+B+\tilde{\Sigma}(E)+i\Gamma_{Z_b}/2}\nonumber\\
&&\times\varepsilon^{ijk}q^i\epsilon^{\ast
j}(h_b)\epsilon^m(\Upsilon).
\end{eqnarray}
E is the energy defined relative to the $BB^\ast$ threshold. B is
the binding energy. $g$ is defined in Eq.~(\ref{gz}). $E_\pi$ is the
energy of $\pi^-$, $p$ is the three-momentum of the $\pi^-$, and $q$
is the three-momentum of the $\pi^+$. $\mu=\frac{m_B
m_{B^\ast}}{m_B+m_{B^\ast}}$ is the reduced mass. Note that the
terms proportional to $p^k p^m$ in $\mathcal{M}_{3c}$ and
$\mathcal{M}_{3d}$ will disappear in the heavy-quark limit, i.e.,
$m_B=m_{B^\ast}$. This indicates that in the heavy-quark limit, the
D wave decay of $\Upsilon(5S)\rightarrow Z_b \pi$ is forbidden. We
neglect the terms proportional to $p^k p^m$ in the fit, since they
will be suppressed by the heavy-quark spin symmetry.

The amplitudes for $\Upsilon(5S)\rightarrow Z_b^{\prime +}
\pi^-\rightarrow h_b(mP) \pi^+\pi^-$ in Fig.~\ref{shortfig}
and~\ref{longfig} read
\begin{eqnarray}
i\mathcal{M}_{2a}&=&\frac{2\sqrt{2Z}g^\prime g_\Upsilon  g_\pi g_h
E_\pi}{F_\pi^2}\frac{1}{E+B^\prime+\tilde{\Sigma}(E)+i\Gamma_{Z_b^\prime}/2}\varepsilon^{ijk}q^i
\epsilon^{\ast
j}(h_b)\epsilon^k(\Upsilon)\nonumber\\
&&\times[I(m_{B^\ast},m_{B^\ast},m_{B^\ast},q)+I(m_{B^\ast},m_{B^\ast},m_B,q)].
\end{eqnarray}

\begin{eqnarray}
i\mathcal{M}_{2b}&=&-i\frac{g_\Upsilon g_z
E_\pi}{F_\pi^2}\frac{Z}{E+B^\prime+\tilde{\Sigma}(E)+i\Gamma_{Z_b^\prime}/2}\varepsilon^{ijk}q^i
\epsilon^{\ast j}(h_b)\epsilon^k(\Upsilon).
\end{eqnarray}

\begin{eqnarray}
i\mathcal{M}_{3a}&=&\frac{2g^{\prime2}g_1 g_\pi g_h
E_\pi}{F_\pi^2}\frac{\mu}{\pi}(-2\mu E-i\epsilon)^{1/2}\frac{1}{E+B^\prime+\tilde{\Sigma}(E)+i\Gamma_{Z_b^\prime}/2}\nonumber\\
&&\times \varepsilon^{ijk}q^i \epsilon^{\ast
j}(h_b)\epsilon^k(\Upsilon)[I(m_{B^\ast},m_{B^\ast},m_{B^\ast},q)+I(m_{B^\ast},m_{B^\ast},m_B,q)].
\end{eqnarray}

\begin{eqnarray}
i\mathcal{M}_{3b}&=&-i\frac{\sqrt{2Z}}{2\pi}\frac{g^\prime g_1 g_z
E_\pi}{F_\pi^2}\mu(-2\mu
E-i\epsilon)^{1/2}\frac{1}{E+B^\prime+\tilde{\Sigma}(E)+i\Gamma_{Z_b^\prime}/2}\varepsilon^{ijk}q^i
\epsilon^{\ast j}(h_b)\epsilon^k(\Upsilon).
\end{eqnarray}

\begin{eqnarray}
i\mathcal{M}_{3c}&=&16\frac{g^{\prime 2} g_2 g_h
g_\pi^2}{F_\pi^2}\left[\left(I^{(1)}(m_{B^\ast},m_{B^\ast},m_{B^\ast},p)+I^{(1)}(m_B,m_{B^\ast},m_{B^\ast},p)\right)\vec{p}^2\delta^{km}+\right. \nonumber\\
&&\left.
\left(I^{(1)}(m_{B^\ast},m_{B^\ast},m_{B^\ast},p)-I^{(1)}(m_{B},m_{B^\ast},m_{B^\ast},p)\right)p^k
p^m\right]\times
\frac{1}{E+B^\prime+\tilde{\Sigma}(E)+i\Gamma_{Z_b^\prime}/2}\nonumber\\
&&\times\varepsilon^{ijk}q^i\epsilon^{\ast
j}(h_b)\epsilon^m(\Upsilon)[I(m_{B^\ast},m_{B^\ast},m_{B^\ast},q)+I(m_{B^\ast},m_{B^\ast},m_{B},q)].
\end{eqnarray}

\begin{eqnarray}
i\mathcal{M}_{3d}&=&-i4\sqrt{2}\frac{g^\prime g_2 g_z g_\pi}{F_\pi^2}\left[\left(I^{(1)}(m_{B^\ast},m_{B^\ast},m_{B^\ast},p)+I^{(1)}(m_{B},m_{B^\ast},m_{B^\ast},p)\right)\vec{p}^2\delta^{km}+\right. \nonumber\\
&&\left.
\left(I^{(1)}(m_{B^\ast},m_{B^\ast},m_{B^\ast},p)-I^{(1)}(m_{B},m_{B^\ast},m_{B^\ast},p)\right)p^k
p^m\right]\times
\frac{\sqrt{Z}}{E+B^\prime+\tilde{\Sigma}(E)+i\Gamma_{Z_b^\prime}/2}\nonumber\\
&&\times\varepsilon^{ijk}q^i\epsilon^{\ast
j}(h_b)\epsilon^m(\Upsilon).
\end{eqnarray}
E is the energy defined relative to the $B^\ast B^\ast$ threshold.
$B^\prime$ is the binding energy. $g^\prime$ is the renormalized
coupling constant which is defined in Eq. (\ref{gz}). We use
$g^\prime$ here to distinguish from $g$, which is used in
$\Upsilon(5S)\rightarrow Z_b^+ \pi^-\rightarrow h_b(mP) \pi^+\pi^-$,
since they may be different due to different binding energies. $\mu$
is the reduced mass of the $B^\ast B^\ast$ system. Other notations
are the same with that in $\Upsilon(5S)\rightarrow Z_b^+
\pi^-\rightarrow h_b(mP) \pi^+\pi^-$. We also neglect the terms
proportional to $p^k p^m$ in the fit.

In the above, we have assumed that $Z_b$ only couples to $B\bar
B^\ast$ while $Z_b^\prime$ only couples to $B^\ast \bar B^\ast$. We
also assume that the probability of finding an elementary state in
$Z_b$ and $Z_b^\prime$ is the same. In other words, we use the same
$Z$ for $Z_b$ and $Z_b^\prime$.

\section*{APPENDIX C: AMPLITUDES FOR $\Upsilon(5S)\rightarrow B^{(\ast)}\bar B^{(\ast)}\pi$}

The amplitudes for $\Upsilon(5S)\rightarrow B^{+}\bar B^{\ast 0}\pi$
in Fig.~\ref{feyndiagram} and~\ref{backfig2} read
\begin{eqnarray}
i\mathcal{M}_{5a}&=&-\frac{g_{\Upsilon} g
E_\pi}{\sqrt{2}F_\pi}\frac{\sqrt{Z}}{E+B+\tilde{\Sigma}(E)+i\Gamma_{Z_b}/2}\epsilon(\Upsilon)\cdot\epsilon^\ast(\bar
B^{\ast 0}).
\end{eqnarray}

\begin{eqnarray}
i\mathcal{M}_{5b}&=&g_1\frac{E_\pi\mu}{2\pi
F_\pi}\frac{g^2\sqrt{-2\mu
E-i\epsilon}}{E+B+\tilde{\Sigma}(E)+i\Gamma_{Z_b}/2}\epsilon(\Upsilon)\cdot\epsilon^\ast(\bar
B^{\ast 0}).
\end{eqnarray}

\begin{eqnarray}
i\mathcal{M}_{5c}&=&4\frac{g^2 g_2 g_\pi}{F_\pi}\epsilon^{\ast j}(\bar B^{\ast 0})\epsilon^m(\Upsilon)\left[\left(I^{(1)}(m_{B^\ast},m_{B^\ast},m_B,q_\pi)+I^{(1)}(m_{B^\ast},m_B,m_{B^\ast},q_\pi)\right)\vec{q}_{\pi}^2\delta^{jm}+\right. \nonumber\\
&&\left.
\left(I^{(1)}(m_{B},m_{B},m_{B^\ast},q_\pi)-I^{(1)}(m_{B^\ast},m_B,m_{B^\ast},q_\pi)\right)q_\pi^j
q_\pi^m\right]\times
\frac{1}{E+B+\tilde{\Sigma}(E)+i\Gamma_{Z_b}/2}.
\end{eqnarray}

\begin{eqnarray}
i\mathcal{M}_{6a}&=&g_1\frac{E_\pi}{F_\pi}\epsilon(\Upsilon)\cdot\epsilon^\ast(\bar
B^{\ast 0}).
\end{eqnarray}

\begin{eqnarray}
i\mathcal{M}_{6b}&=&-4\frac{g_2
g_\pi}{F_\pi}\epsilon^i(\Upsilon)\epsilon^{\ast j}(\bar B^{\ast
0})\left[\frac{1}{E_\pi+\Delta}p_B^i
q_\pi^j-\frac{1}{E_\pi}(q_{\pi}^i p_B^j-q_\pi\cdot
p_B\delta^{ij})+\frac{1}{E_\pi-\Delta}(q_\pi^i p_{\bar
B}^j+q_\pi\cdot p_{\bar B}\delta^{ij}-q_\pi^j p_{\bar
B}^i)\right].\nonumber\\
\end{eqnarray}
E is the energy defined relative to the $BB^\ast$ threshold. $E_\pi$
is the pion energy, $q_\pi$ is the three-momentum of the pion. $\mu$
is the reduce mass of $BB^\ast$ system. $p_B$ and $p_{\bar B}$ are
the three-momentum of $B^+$ and $\bar B^{\ast 0}$, respectively.
$\Delta$ is the
hyperfine splitting of the B mesons.\\

The amplitudes for $\Upsilon(5S)\rightarrow  B^{\ast +} \bar B^{\ast
0}\pi$ in Fig.~\ref{feyndiagram} and Fig.~\ref{backfig2} read
\begin{eqnarray}
i\mathcal{M}_{5a}&=&\frac{g_{\Upsilon} g^\prime
E_\pi}{\sqrt{2}F_\pi}\frac{i\sqrt{Z}}{E+B^\prime+\tilde{\Sigma}(E)+i\Gamma_{Z_b^\prime}/2}
\varepsilon^{imn}\epsilon^i(\Upsilon)\epsilon^{\ast m}(\bar B^{\ast
0})\epsilon^{\ast n}(B^{\ast +}).
\end{eqnarray}

\begin{eqnarray}
i\mathcal{M}_{5b}&=&ig_1\frac{E_\pi\mu}{2\pi F_\pi}\frac{g^{\prime
2}\sqrt{-2\mu
E-i\epsilon}}{E+B^\prime+\tilde{\Sigma}(E)+i\Gamma_{Z_b^\prime}/2}\varepsilon^{imn}\epsilon^{i}(\Upsilon)\epsilon^{\ast
m}(\bar B^{\ast 0})\epsilon^{\ast n}(B^{\ast +}).
\end{eqnarray}

\begin{eqnarray}
i\mathcal{M}_{5c}&=&i4\frac{g^{\prime 2} g_2 g_\pi}{F_\pi}\left[\left(I^{(1)}(m_{B^\ast},m_{B^\ast},m_{B^\ast},q_\pi)+I^{(1)}(m_{B},m_{B^\ast},m_{B^\ast},q_\pi)\right)\vec{q}_\pi^2\delta^{km}+\right. \nonumber\\
&&\left.
\left(I^{(1)}(m_{B^\ast},m_{B^\ast},m_{B^\ast},q_\pi)-I^{(1)}(m_{B},m_{B^\ast},m_{B^\ast},q_\pi)\right)q_\pi^k
q_\pi^m\right]\times
\frac{1}{E+B^\prime+\tilde{\Sigma}(E)+i\Gamma_{Z_b^\prime}/2}\nonumber\\
&&\times\varepsilon^{kji}\epsilon^{m}(\Upsilon)\epsilon^{\ast
j}(\bar B^{\ast 0})\epsilon^{\ast i}(B^{\ast +}).
\end{eqnarray}

\begin{eqnarray}
i\mathcal{M}_{6a}&=&-ig_1\frac{E_\pi}{F_\pi}\varepsilon^{imn}\epsilon^i(\Upsilon)
\epsilon^{\ast m}(B^{\ast +})\epsilon^{\ast n}(\bar B^{\ast0}).
\end{eqnarray}

\begin{eqnarray}
i\mathcal{M}_{6b}&=&-i4\frac{g_2
g_\pi}{F_\pi}\epsilon^{i}(\Upsilon)\epsilon^{\ast j}(B^{\ast
+})\epsilon^{\ast k}(\bar B^{\ast
0})\left[\frac{1}{E_\pi+\Delta}\left(-\varepsilon^{ijm}p_B^m
q_\pi^k+\varepsilon^{ikm}p_{\bar B}^m q_\pi^j\right)\right.\nonumber\\
&&\left.+\frac{1}{E_\pi}\left(\varepsilon^{jnm}\delta^{ik}p_{\bar
B}^m p_B^n-\varepsilon^{kmn}\delta^{ij}p_{\bar B}^m
p_B^n+\varepsilon^{mjk}q_\pi^m q_\pi^i-\varepsilon^{mki}q_\pi^m
p_B^j+\varepsilon^{mji}q_\pi^m p_{\bar B}^k\right)\right].
\end{eqnarray}
E is the energy defined relative to the $B^\ast B^\ast$ threshold.
$E_\pi$ is the pion energy, $q_\pi$ is the three-momentum of pion.
$\mu$ is the reduce mass of $B^\ast B^\ast$ system. $p_B$ and
$p_{\bar B}$ are the three-momentum of $B^{\ast +}$ and $\bar
B^{\ast 0}$, respectively. $\Delta$ is the hyperfine splitting of
the B mesons.

\end{document}